\documentclass[aps,prl,showpacs,twocolumn,floats,superscriptaddress,floatfix]{revtex4-1}
\usepackage{amsmath,amssymb}
\usepackage{graphicx}
\usepackage{psfrag}
\usepackage{graphicx}

\begin{document}

\title{Universality of scaling and multiscaling in turbulent symmetric binary fluids}
\author{Samriddhi Sankar Ray}
\email{samriddhisankarray@gmail.com}
\affiliation{Laboratoire Cassiop\'ee, Observatoire de la C\^ote
d'Azur, UNS, CNRS, BP 4229, 06304 Nice Cedex 4, France.}
\author{Abhik Basu}
\email{abhik.basu@saha.ac.in}
\affiliation{Theoretical Condensed Matter Physics Division, Saha Institute of Nuclear Physics,
1/AF Bidhannagar, Kolkata (Calcutta) 700064, India}
\begin{abstract}
We elucidate the universal scaling and multiscaling properties of
the nonequilibrium steady states (NESS) in a driven symmetric
binary fluid (SBF) mixture in its homogeneous miscible phase in
three dimensions ($3d$). We show, for the first time,
via Direct Numerical Simulations (DNS)
that structure functions of the velocity and the concentration gradient
exhibit multiscaling in $3d$ and extended
self-similarity (ESS). We also find that, in contrast to the
well-known passive scalar turbulence problem, structure functions of the
concentration show simple scaling. We propose a new shell model
for SBF turbulence which preserve all the invariances in the
ideal limit of the SBF equations and which reduces to a well-known
shell model for fluid turbulence in the zero concentration field limit.
We show that the shell model has the same scaling properties as
the $3d$ SBF equations.
Our combined results from our DNS of the SBF equations and
shell-model studies consistently bring out the multiscaling of the
velocity and concentration gradient fields and simple scaling of the
concentration field.
\end{abstract}
\maketitle

\section{Introduction}

The scaling properties of correlation functions near a critical
point in equilibrium statistical mechanics have been well
understood over the past few decades. However, understanding
similar power-law
scaling behaviours in structure functions in a variety of
turbulent flows remains an open problem
in nonequilibrium statistical mechanics ~\cite{frisch}.
In recent years, significant progress has been made in the
study of {\it equal-time} structure functions in the
turbulence of fluids, magnetohydrodynamics (MHD) and, most
notably, passive-scalars~\cite{falcormp}. By contrast,
for symmetric binary fluid turbulence, statistical studies are
still in its infancy.

To appreciate the context and the necessity for a systematic study
of the scaling properties of equal-time structure functions in such
symmetric binary fluid mixtures, it is important to recall some
lessons from standard equilibrium critical
phenomena~\cite{rmp,chaikin}. For a $d$ dimensional spin system near
a critical point, the equal-time, correlation function $g$, between
two spins separated by the vector ${\bf r}$ ($r = |{\bf r}|$), and
its spatial Fourier transform $\tilde g$ have power-law scalings :
\begin{eqnarray}
g({\bf r}; \bar{t},h) &\approx&
\frac{G(r\bar{t}^{\nu},h/\bar{t}^{\Delta})}{r^{d-2+\eta}}\/;
\nonumber \\
\tilde g({\bf k}; \bar{t},h)  &\approx&
\frac{{\tilde G}(k/\bar{t}^{\nu},h/\bar{t}^{\Delta})}{k^{2-\eta}}
 \/,
\end{eqnarray}
where, $\bar{t}\equiv (|T - T_c|)/T_c$, $T$ and $T_c$ are the
temperature and the critical temperature, respectively,
$h \equiv H/k_BT_c$, $H$ is the external field, $k_B$ is the
Boltzmann constant, $\bf{k}$ is the wavevector,
$k=|{\bf k}|$, $\nu$, $\Delta$, $\eta$ are critical exponents, and
$G$ and ${\tilde G}$ are scaling functions.  We note, in passing,
that away from the
critical point the correlation functions decay exponentially and the
associated correlation length $\xi_c$ diverges near a
critical point as $\xi_c \sim \bar{t}^{-\nu}$, if $h = 0$.

Can we generalise such ideas of equilibrium statistical mechanics
to the case of homogeneous, isotropic  turbulence in various settings?
Indeed, the power-law behaviours of equal-time structure functions, in the inertial
range (to be defined later), in fluid, passive-scalar or MHD turbulence
have a certain similarity to the algebraic dependence on $r$ of correlation functions in
critical theory. To make this connection explicit and lay the ground
for our subsequent discussions, we begin with the increments of
the longitudinal component of the velocity
$\delta u_{\parallel}({\bf x},{\bf r},t) \equiv [{\bf u}({\bf x}+
{\bf r},t) - {\bf u}({\bf x},t)]\cdot({\bf r}/r)$, where
${\bf u}({\bf x},t)$ is
the velocity of the fluid at the
point ${\bf x}$ and time $t$, and the subscript $\parallel$ implies the
longitudinal component. The order-$p$, equal-time structure
functions for the fluid (superscript $u$)  field are defined conventionally as
\begin{equation}
{\cal S}^{u}_p(r) \equiv \langle [\delta u_{\parallel}
({\bf x},{\bf r},t)]^p \rangle
\sim r^{\zeta^{u}_p};
\label{Sp}
\end{equation}
the angular brackets indicate averages over the steady state for
statistically steady turbulence or over statistically independent
initial configurations for decaying turbulence.
The power law behaviour of such structure functions, which is valid
for separations $r$ in the inertial range  $\eta_d \ll r \ll L$, where
$\eta_d$ is the Kolmogorov dissipation scale and $L$ the large
length scale at which energy is injected into the system are
characterised by the equal-time exponents $\zeta^{u}_p$.

Kolmogorov's phenomenological theory \cite{frisch,K41a, K41b} of
1941 (K41) for fluid turbulence, predicts simple scaling $\zeta^{u,K41}_p = p/3$.
Subsequent experimental and numerical studies, however, strongly suggests the
existence of equal-time multiscaling : $\zeta^{u}_p$ is a nonlinear,
convex, monotone-increasing functions of $p$.
Indeed, it is important to remember that for the simplified stochastic
Kraichnan model \cite{falcormp,kraich1,kraich2,kraich3}
of passive-scalar turbulence, multiscaling of equal-time structure functions
can be demonstrated analytically. The analogue
of the K41 theory for passive-scalar turbulence is due to Obukhov
and Corrsin \cite{obu,corr}. For the Schmidt number
$Sc \equiv \nu/\kappa \simeq 1$, where $\nu$ is the kinematic
viscosity of the fluid and $\kappa$ is the diffusivity of the
passive scalar, the Obukhov-Corrsin theory leads to K41
scaling exponents for the passive-scalar case.

In sharp contrast to fluid and passive-scalar turbulence, a systematic theoretical
and numerical study
of the statistical properties of symmetric, binary fluid (SBF)
mixtures in three dimensions ($3d$)
is still in its early stages and experiments performed on such
systems have been typically concerned with measurements of
effective transport coefficients \cite{exp}. Our prime concern here is to extend
the ideas of equal-time scaling and multiscaling to the turbulence of
SBF. In this paper we
provide for the first time, via detailed Direct Numerical
Simulations (DNS) and a new shell model that we propose for
such a system, a systematic study of the statistical
properties of equal-time, two-point structure functions in
a statistically steady, turbulent SBF mixture.
We thus consider an incompressible, binary fluid mixture,
with components labelled $A$ and $B$, and having densities
$\rho_A({\bf x},t)$ and $\rho_B({\bf x},t)$, respectively, such
that the concentration field $\psi ({\bf x},t)$ is defined
via $\psi ({\bf x},t) \equiv
[\rho_A({\bf x},t)-\rho_B({\bf x},t)]/\rho_0$, where $\rho_0$ is
the mean density. Furthermore, since we will be interested in
a {\em symmetric}, binary fluid mixture we impose the constraint
$\langle \psi({\bf x},t)\rangle =0$.
We elucidate the universal properties of homogeneous, isotropic SBF
turbulence, in the absence of any macroscopic (mean) concentration
gradient, by measuring the scaling exponents of the equal-time
structure functions of the velocity field $\bf u$, the concentration
field $\psi$ and the concentration gradient field ${\bf
b}=\nabla\psi$. We show for the first time that although the
exponents associated with $\bf b$ are multiscaling (like the
exponents for $\bf u$), the equal-time exponents for $\psi$ show
simple-scaling. Our results are similar to the numerical
quasi-Lagrangian (in two-dimensional flows) \cite{celani} and in
agreement with the predictions of one-loop field theoretical
\cite{abhik} studies of the SBF system.


\section{Model equations}

In order to describe the coupled dynamical evolution of the field $\bf
u$ and $\psi$ we need coupled dynamical equations for $\bf u$ and
$\psi$. The equation of motion of the velocity $\bf u$ is the
generalised Navier-Stokes equation which now includes the stresses
from the $\bf b$ field \cite{ruiz,jkb1}
\begin{equation}
\frac{\partial {\bf u}}{\partial t}+\lambda_1({\bf u}\cdot
\nabla){\bf u}= -\frac{\nabla P}{\rho_0}- \lambda_2 \nabla\psi
\nabla^2\psi +\nu\nabla^2 {\bf u} +{\bf f},\label{navierI}
\end{equation}
where $\lambda_1,\,\lambda_2 >0$ are coupling constants, and
advection diffusion equation for $\psi$:
\begin{equation}
\frac{\partial \psi}{\partial t} + \lambda_3{\bf u}\cdot \nabla\psi
= \eta\nabla^2\psi + f_\psi.\label{advecI}
\end{equation}
In Eqs.~\ref{navierI} and \ref{advecI}, $P$ and $\rho_0$ are the
local (effective) pressure and density, respectively; since we
consider an incompressible fluid we further have $\rho_0={\rm
const.}$ and $\nabla\cdot{\bf u}=0$. The constants $\nu$ and $\eta$
are the kinematic viscosity and concentration diffusivity,
respectively. The functions $\bf f$ and $f_\psi$ are forcing terms
which drive the system to a statistically steady state. Galilean
invariance of the system enforces $\lambda_1=\lambda_3=1$
\cite{jkb1,abhik}. Further, $\lambda_2$ may be set to unity by
appropriately choosing the unit of $\psi$ (equivalently, by
exploiting the rescaling invariance of $\psi$)\cite{abhik}. Thus, in
what follows, we set $\lambda_1=\lambda_2=\lambda_3=1$.  It is clear
from Eqs.~(\ref{navierI}) and (\ref{advecI}) (see also
Refs.~\cite{celani,ruiz,jkb1}) that in the dynamics of a symmetric
binary fluid mixture, the velocity field $\bf u$ couples with the
concentration gradient ${\bf b} =\nabla\psi$ and not with $\psi$
itself. Thus it is useful to write the coupled evolution equations
in terms of $\bf u$ and $\bf b$: The resulting equations are
\begin{equation}
\frac{\partial {\bf u}}{\partial t}+({\bf u}\cdot \nabla){\bf u}=
-\frac{\nabla P}{\rho_0}-{\bf b}\nabla\cdot {\bf b} +\nu\nabla^2
{\bf u} +{\bf f} \label{navier}
\end{equation}
and the advection-diffusion equation for $\bf b$
 is \cite{ruiz,jkb1}
\begin{equation}
\frac{\partial {\bf b}}{\partial t}+\nabla\cdot ({\bf u\cdot
b})=\eta\nabla^2 \bf b+{\bf g}. \label{advec}
\end{equation}
Here, ${\bf g}=\nabla f_\psi$. Note that $\nabla\times {\bf b}=0$,
thus $\bf b$ is an {\em irrotational} field. In a symmetric binary
mixture, $\psi$ is not advected {\it passively} by the velocity
field, but is {\em active}, i.e., the concentration gradient ${\bf
b}$ reacts back on $\bf u$ and thus modifies the flow. Furthermore,
since we are interested in the isotropic and homogeneous case, i.e.,
we have no mean concentration gradient, we impose $\langle \bf b
\rangle=0$.



\section{STRUCTURE FUNCTIONS AND MULTISCALING}

The order-$p$, equal-time structure function is defined as ${\cal
S}_p^a (r)=\langle |a({\bf x+r}) - a({\bf x})|^p\rangle$, where $a$
can be ${\bf u,\,b}$ or $\psi$, $\bf x,\,r$ are spatial coordinates
and the angular brackets represent an average over the NESS. For
${\bf r}$ in the inertial range which lies between the large length scale $L$ and
$\eta_d$, the Kolmogorov scale where dissipation becomes
significant, and at high fluid and concentration-gradient Reynolds
numbers, $Re$ and ${\rm Re_b}$, respectively, we
expect power-law scaling ${\cal S}_p^a (r)\sim r^{\zeta_p^a}$.
 The determination of the exponents $\zeta_p^a$ has been
one of the central, but still elusive, goals of studies in the
statistical theory of turbulence.
The extension of Kolmogorov's 1941 theory
\cite{K41a} to homogeneous, isotropic SBF turbulence, with no mean
concentration gradient, yields $\zeta_p^a=p/3$, i.e., {\em simple
scaling}. In isotropic and homogeneous pure fluid 
turbulence,  we have corrections to simple-scaling exponents such
that the equal-time exponents for such systems $\zeta_p^u =p/3
-\delta\zeta_p^u$, where $\delta\zeta_p^u > 0$ and $\zeta_p^u$ is a
nonlinear, monotonically increasing functions of $p$. Extensive
analytical and numerical studies on the well-known passive scalar
problem~\cite{passive}, which is the passive limit of the system
considered here, clearly demonstrate that $\zeta_p^\psi$ has
multiscaling qualitatively similar to $\zeta_p^u$. In contrast, it
has been shown for two-dimensional flows in Ref.~\cite{celani}, by
using a Lagrangian approach, that $\zeta_p^\psi = p/3$, i.e., ${\cal
S}_p^\psi (r)$ shows only simple scaling. Ref.~\cite{abhik} used
symmetry arguments to show that $\zeta_p^\psi=p/3$ and suggested
that $\zeta_p^b$ should show multiscaling akin to $\zeta_p^u$. It is
thus expected that the multiscaling behaviour of the NESS in
homogeneous and isotropic SBF turbulence is characterised by
$\zeta_p^u$ and $\zeta_p^b$. Here, we confirm this in numerical
studies of $3d$SBF equations and our shell model equations.

Before we embark upon a discussion of our results, we explore
the formal similarities between the dynamical equations of binary
fluid turbulence and MHD. These become apparent when Eqs.
(\ref{navierI}) and (\ref{advecI}) are compared with the
incompressible $3d$MHD equations.
%
%
The incompressible $3d$MHD equations
are given by \cite{mhd-basic}
\begin{eqnarray}
\frac{\partial {\bf u}}{\partial t}+({\bf u}\cdot \nabla){\bf u} &=&
-\frac{\nabla P}{\rho_0} + \frac{\bf (\nabla\times B)\times
B}{4\pi\rho_0}  +\nu\nabla^2 {\bf u} +{\bf f}; \nonumber \\
\frac{\partial {\bf B}}{\partial t} +\nabla\times ({\bf u\times
B}) &=& \mu_0\nabla^2 {\bf B} + \bf g. \label{3dmhd}
\end{eqnarray}
Here $\bf B$  and $\mu_0$ are the magnetic field and magnetic
viscosity, respectively. Other symbols have the same meaning as in
Eqs.~(\ref{navier}) and (\ref{advec}). The similarities between
Eqs.~(\ref{navier}), (\ref{advec}) and (\ref{3dmhd}) are noteworthy:
(i) The concentration gradient field $\bf b$ and the magnetic field $\bf B$
have the same {\em na\"ive} dimensions (which in turn is the same as the
na\"ive dimensions of the velocity $\bf u$); and (ii) the non-linear terms
in Eqs.~(\ref{navier}) and (\ref{advec}) have analogues (in the
sense of the number of field and gradients) in Eqs.~(\ref{3dmhd})
with same na\"ive dimensions. All these suggest that the
concentration gradient field $\bf b$ plays the role of the magnetic
field in MHD \cite{foot}. In homogeneous and isotropic $3d$MHD
turbulence,  structure functions ${\cal S}_p^a (r)=\langle |a({\bf
x+r}) - a({\bf x})|^p\rangle$ where $a$ refers to both $u$ and $B$,
exhibit multiscaling similar to pure fluid turbulence. This suggests that
the structure functions ${\cal S}_p^b(r)$, like the magnetic field
structure functions in MHD \cite{abprl}, should exhibit multiscaling
akin to pure fluid turbulence. In this paper, we confirm this conjecture.

A promising starting point for a systematic theory is one
where Eqs.\ref{navier} and \ref{advec} are forced by Gaussian random
forces $\bf f$ and $\bf g$ [cf. Refs.~\cite{yakhot} for an
application of this approach in pure fluid turbulence], whose
spatial Fourier transforms, ${\bf f}({\bf k},t)$ and ${\bf g}({\bf
k},t)$, respectively, have zero mean and covariances $\langle f_i
({\bf k},t)f_j({\bf -k},0)\rangle = A_f P_{ij}({\bf
k})k^{4-d-y}\delta(t)$, $\langle  g_i({\bf k},t)g_j({-\bf
k},0)\rangle =A_\psi k_ik_jk^{2-d-y}\delta (t)$ (corresponding to
noise variance $\langle f_\psi({\bf k}, t)f_\psi({\bf
-k},0)\rangle=A_\psi k^{2-d-y}$), where $\bf k$ is a wavenumber, $t$
time, $i,j$ Cartesian components in $d$-dimensions, $A_f$ and
$A_\psi$ are a constant amplitude and $P_{ij}({\bf
k})=\delta_{ij}-k_ik_j/k^2$ is the transverse projector which
enforces the incompressibility condition \cite{abhik}. One-loop
renormalisation group studies \cite{abhik,jkb1} of this model
yield K41 energy spectra for $\bf u$ and $\bf b$ fields: $E^{u,b}
(k)\sim k^2 {\cal S}_2 ^{u,b} (k)\sim k^{-5/3}$ for $d=3$ and $y=4$.
Nevertheless, these RG studies have been criticised for a variety of
reasons \cite{rgprobs} such as using a large value for $y$ in a
small-$y$ expansion and neglecting an infinity of marginal operators
(if $y = 4$). These criticisms of the approximations, however
justified they may be, cannot be used to rule out the randomly forced model as an
appropriate theory for SBF turbulence. In this paper, we use random Gaussian
forcing in all our numerical simulations.


We end this Section by summarising the key results in this paper.
Our studies yield many interesting results: The multiscaling
exponents for $\bf u$ and $\bf b$ fields which we obtain from
$3d$SBF and our shell models agree [Figs.~\ref{DNS_zetap},
~\ref{DNS_zetab}] and $\zeta_p^u\sim\zeta_p^b$ lie close to, but
below, the She-Leveque prediction (SL) \cite{sl} for pure fluids
($\zeta_p^{SL}=p/9 + 2[1-(2/3)^{p/3}]$). Furthermore, the
probability distribution functions (PDF) (Fig.~\ref{PDF}) for
$\delta a_\alpha ({\bf r})= a_\alpha ({\bf x +r}) - a_\alpha({\bf
x}),\,a=u,b$ show non-Gaussian tails, whereas the same for
$\delta\psi ({\bf r})$ shows good agreement with a Gaussian
distribution. These features of the PDF confirm the multiscaling
behaviour of $\bf u,\,b$ and the simple scaling of $\psi$. Earlier studies of
fluid~\cite{fluid-ess} and MHD~\cite{abprl} turbulence show that an
extended inertial range is obtained if we use Extended Self
Similarity (ESS): Thus, by making use of ESS, in which
$\zeta_p^a/\zeta_3^a$ follows from ${\cal S}^a_p \sim [{\cal
S}^a_3]^{\zeta_p^a/\zeta_3^a},\,a=u,b$, and $\psi$, we expect, by
analogy, that the scaling range $r$ extends down to $r\simeq
5\eta_d$. We confirm this in our simulations for SBF turbulence.

\subsection{SHELL MODEL FOR THE SBF MIXTURE}
As is well known in turbulence, it is important to resolve the large
ranges of both temporal and spatial scales well. A DNS approach to
hydrodynamical partial differential equations, such as the one we use for the SBF mixture,
is often very difficult if we want to resolve all the scales relevant to turbulence.
To gain insight it is thus useful to consider
simplified models of turbulence that are numerically more tractable than
the partial differential equations themselves.
{\it Shell models} are important examples of such simplified
models; they have proved particularly useful for testing ideas of multiscaling
in fluid, passive-scalar and MHD turbulence~\cite{frisch,abprl,goy,raynjp}. Keeping this in mind, we
derive below a new shell model for the gradient of the concentration field in the SBF system and solve it numerically
to obtain results in support of our DNS results. We should point out that for SBF turbulence,
a shell model was derived in Ref.~\cite{jensen} for the scalar concentration field and its coupling
with the fluid field.

Shell models cannot be derived from the hydrodynamical equations in any rigorous way.
Such models are constructed on a basis of a discretised Fourier space with logarithmically
spaced wave vectors $k_n= k_0 {\tilde \lambda}^n\/, {\tilde \lambda} > 1$ which are associated with
shells $n$ and dynamical complex, scalar dynamical variables which mimic, e.g., velocity increments $u_n$
over scales $\propto 1/k_n$. Furthermore, we impose that $k_n$ be a scalar because spherical symmetry is
implicit in Gledzer-Ohkitani-Yamada (GOY)-type shell models which study homogeneous, isotropic turbulence~\cite{goy,raynjp}.
The logarithmic discretisation of the Fourier space allows
us to reach very high Reynolds number (which are impossible using DNS in present-day computers)
even with moderate values of  $N$, where $N$ is the total number of shells.

The temporal evolution of such a shell model is governed by a set of
ordinary differential equations
that have certain features in common with the Fourier-space version of
the hydrodynamical  equation. Thus, the shell model analogue of the Navier-Stokes equation, for example,
will have a viscous-dissipation term  of the form $-\nu k_n^2u_n$ and
nonlinear terms of the form $\imath k_nu_nu_{n'}$ that couple
velocities in different shells. (We note in passing that gradients appear as products of
$k_n$ in shell models.) In the Navier-Stokes equation all Fourier modes of
the velocity are coupled to each other directly but in most shell models nonlinear interactions are
limited to shell velocities in nearest-  and
next-nearest-neighbour shells. Hence {\it sweeping effects} common to equations of
hydrodynamics, are absent in shell models.

Keeping in mind the constraints of the hydrodynamical equations themselves,
we propose the evolution equations for the shell model analogues of the velocity
$u_n$, the concentration field $\psi_n$ (see also Ref.~\cite{jensen}), and the gradient of the
concentration field $b_n$ as
\begin{eqnarray}
[\frac{d}{dt} + \nu k_n^2]u_n &=&  i[A_1 k_n u_{n+1}u_{n+2} +
A_2 k_{n-1} u_{n-1}u_{n+1} \nonumber \\ &+& A_3 k_{n-2} u_{n-1}u_{n-2} +
A_4 k_n b_{n+1}b_{n+2} \nonumber \\ &+& A_5 k_n b_{n+1}b_{n-1} + A_6 k_{n_2}b_{n-1}b_{n-2}]^{\ast} \nonumber \\ &+& f_n,
\label{goy}
\end{eqnarray}

\begin{eqnarray}
[\frac{d}{dt} + \eta k_n^2]\psi_n &=&  i[k_n(\psi_{n+1}u_{n-1} - \psi{n-1}u_{n+1}) \nonumber \\ &-& \frac{k_{n-1}}{2}(\psi_{n-1}u_{n-2} + \psi_{n-2}u_{n-1}) \nonumber \\ &-& \frac{k_{n+1}}{2}(\psi_{n+2}u_{n+1} + \psi_{n+1}u_{n+2})]^{\ast} \nonumber \\ &+& g_n,
\label{psi}
\end{eqnarray}
and
\begin{eqnarray}
[\frac{d}{dt} + \eta k_n^2]b_n &=&  i[A_7 k_n( u_{n+1}b_{n+2} +
u_{n+2}b_{n+1}) \nonumber \\ &+& A_8 k_{n-1}(u_{n+1}b_{n-1} +
u_{n-1}b_{n+1} \nonumber \\ &+& A_9 k_{n-2}(u_{n-1}b_{n-2} + u_{n-2}b_{n-1})]^{\ast} \nonumber \\ &+& g_n,
\label{b}
\end{eqnarray}
respectively.
In these equations, complex conjugation is denoted by $\ast$,
and the coefficients are chosen such that the shell model
analogues of total energy and the total autocorrelation of the
concentration field is conserved in the absence of forcing
and dissipation. Thus we obtain
$A_1 = 1$, $A_2 = \epsilon - 1$, $A_3 = \epsilon$,
$A_4 + A_8 + A_9 = 0$, $A_5 - A_7 + A_9 = 0$, $A_6 + A_7 + A_8 = 0$,
$A_7 + A_9/\lambda^4 = 0$, $A_7 -A_8/\lambda^2 = 0$, and
$A_8 + A_9/\lambda^2 = 0$. We use the usual GOY model choice~\cite{goy}
of $\epsilon = 0.5$ and fix $A_7 = 1$ in order to obtain the
values of the remaining constants. We have checked that our results
are insensitive to the choice of $A_7$.

\subsection{RESULTS FROM SHELL MODEL AND DNS STUDIES}
\begin{center}
\begin{figure}
\includegraphics[height=6cm,width=6cm]{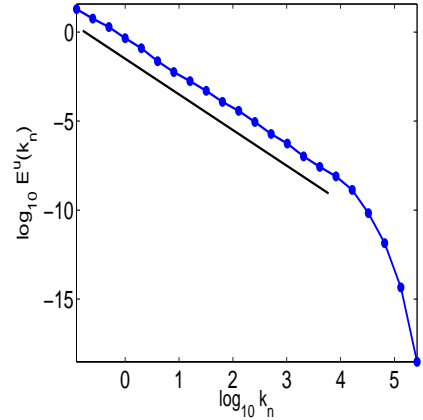}
\caption{(Color online) Log-log plot of the fluid kinetic
energy $E^u(k_n)$ versus the wavenumber $k_n$ (blue filled circle
joined by the continuous line) from our shell model studies.
The thick black line indicates the K41 scaling.}
\label{Ekshell}
\end{figure}
\end{center}
\begin{center}
\begin{figure}
\includegraphics[height=6cm,width=6cm]{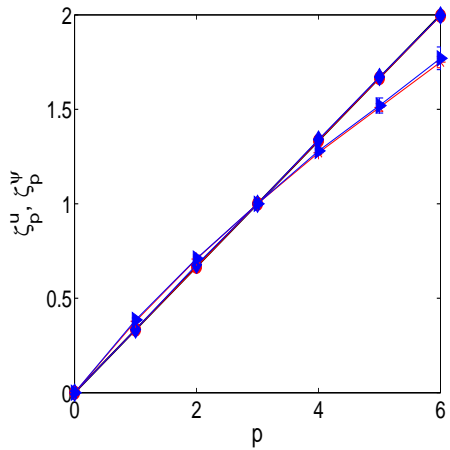}
\caption{(Color online) Plots of  $\zeta_p^u$ with error bars
from our $128^3$ DNS (blue triangle) and shell model (red star),
$\zeta_p^\psi$ with error bars from our $128^3$ DNS (blue diamond)
and shell model
(red filled-circle), and K41 scaling (thick black line)
versus $p$. The lines connecting the data
points from our simulations are a guide to the eye. The data from our
DNS (upto $1 \le p \le 6$) and shell model (shown for $1\le p \le 6$) are
almost indistinguishable from each other upto $p = 6$.}
\label{DNS_zetap}
\end{figure}
\end{center}
\begin{center}
\begin{figure}
\includegraphics[height=6cm,width=6cm]{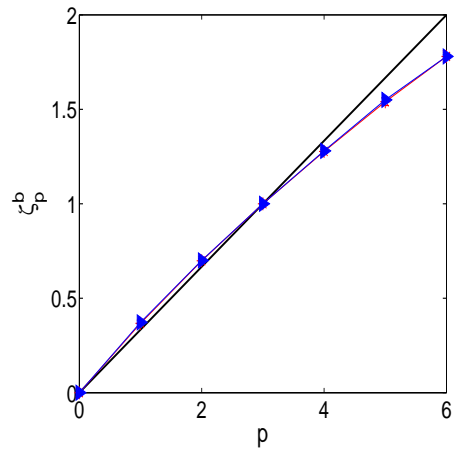}
\caption{(Color online) Plots of  $\zeta_p^b$ with error bars
from our $128^3$ DNS (blue triangle) and shell model (red star), and
K41 scaling (thick black line) versus $p$. The lines connecting the data
points from our simulations are a guide to the eye. The data from our
DNS (upto $1 \le p \le 6$) and shell model (shown for $1\le p \le 6$) are
almost indistinguishable from each other upto $p = 6$.}
\label{DNS_zetab}
\end{figure}
\end{center}
\begin{center}
\begin{figure}
\includegraphics[height=6cm,width=7cm]{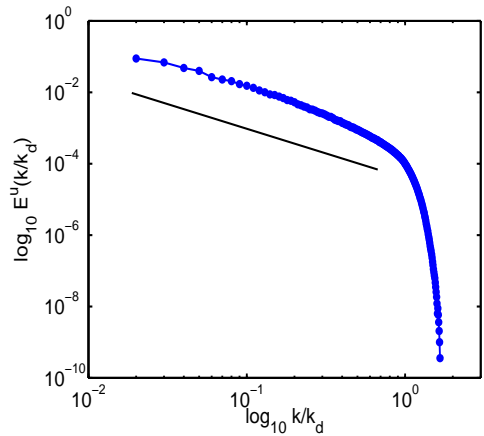}
\caption{(Color online) Log-log plot of $E^u(k)$ versus $k/k_d$ (blue *) from our DNS
studies (see text). $k_d\sim 98 $ is the dissipation scale wavenumber here. The thick black line indicates the K41 scaling.}
\label{Ekdns}
\end{figure}
\end{center}
\begin{center}
\begin{figure}
\includegraphics[height=6cm,width=6cm]{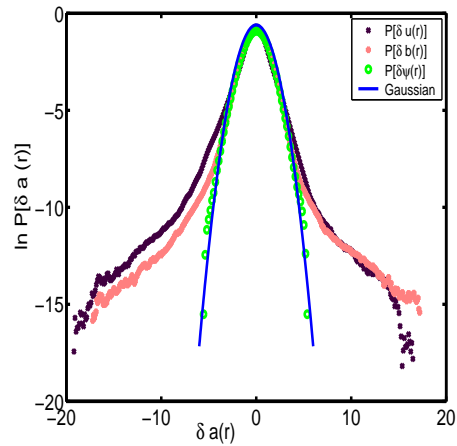}
\caption{(Color online) Semilog  plots of the probability
distributions $P[\delta u (r)],\,P[\delta b (r)]$ and $P[\delta \psi
(r)]$ versus $r$ in the inertial range, averaged over both time and
the Cartesian components, from our DNS; a Gaussian distribution
(blue continuous line) is shown for comparison. } \label{PDF}
\end{figure}
\end{center}
\begin{figure}[htb]
\includegraphics[height=6cm,width=6cm]{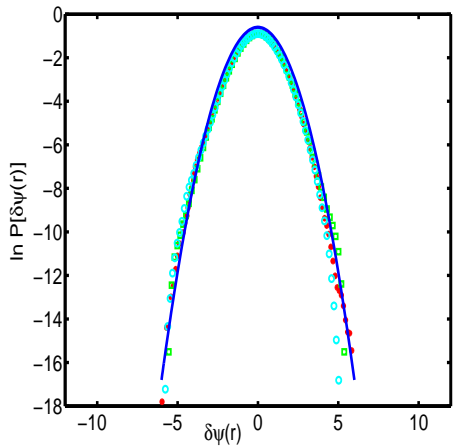}
\caption{(Color online) Semilog plots of normalized probability distributions
$P[\delta\psi(r)]$ as functions of $\delta\psi (r)$ for three
different separations $r$ in the inertial range. A semilog plot of
normalized Gaussian is shown for comparison. All plots are
overlapping (see text).} \label{prob2}
\end{figure}
\begin{figure}[htb]
\includegraphics[height=6cm,width=6cm]{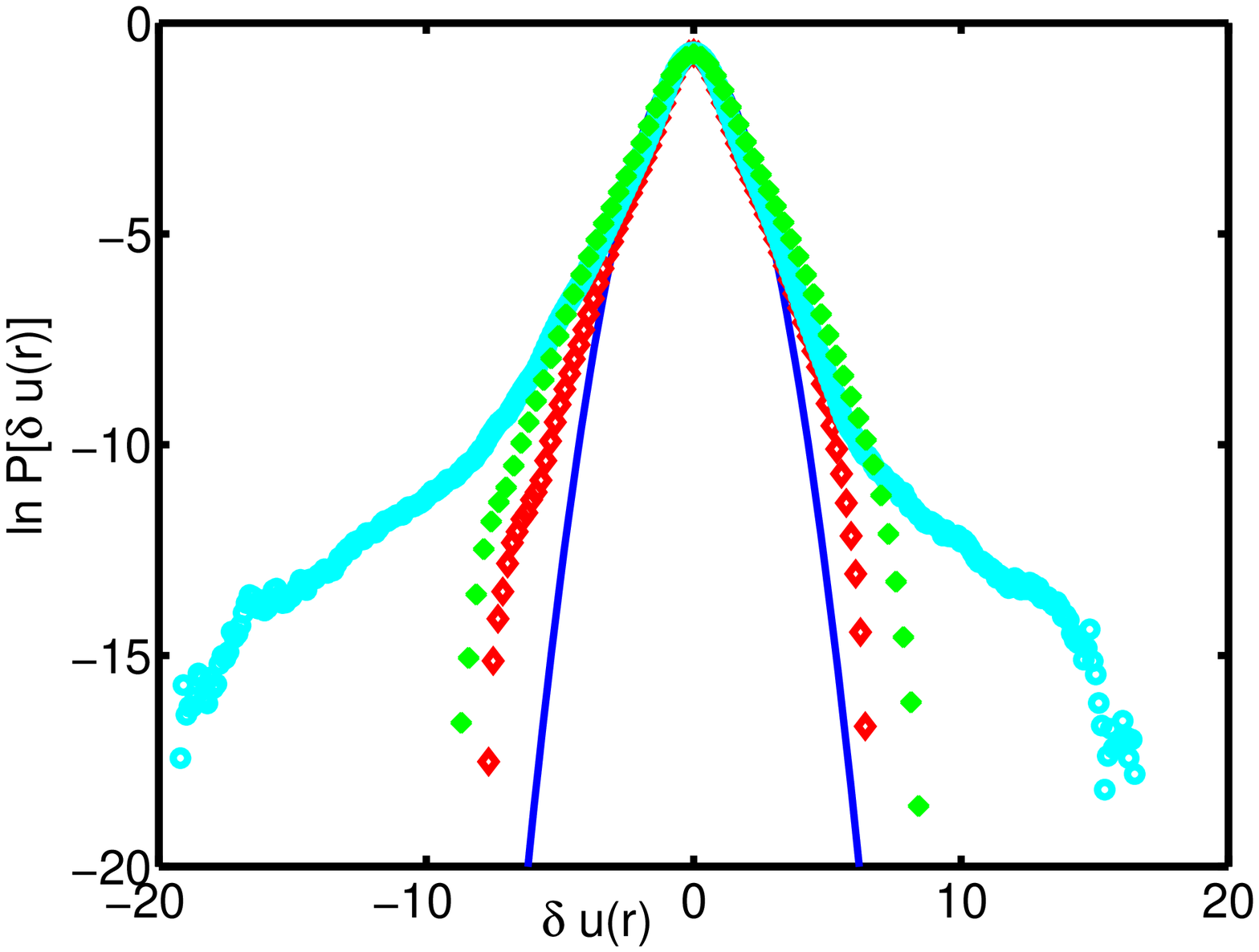}
\caption{(Color online) Semilog plots of normalized probability distributions
$P[\delta u(r)]$ as functions of $\delta u(r)$ for three
different separations $r$ in the inertial range. A semilog plot of
normalized Gaussian is shown for comparison (see text). The values
of $r$ increases going inwards.} \label{prob2u}
\end{figure}
\begin{figure}[htb]
\includegraphics[height=6cm,width=6cm]{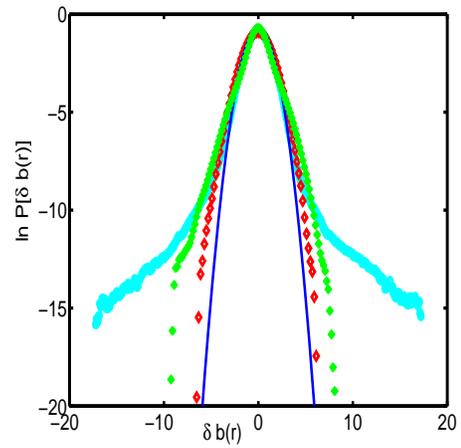}
\caption{(Color online) Semilog plots of normalized probability distributions
$P[\delta b(r)]$ as functions of $\delta b(r)$ for three
different separations $r$ in the inertial range. A semilog plot of
normalized Gaussian is shown for comparison (see text). The values of $r$ increases going inwards.} \label{prob2b}
\end{figure}

We begin by describing results from our numerical simulations of the
shell model for the SBF mixture. In our simulations the shell number
is chosen such that $1 \leq n \leq N$, where $N = 22$ is the total
number of shells and we use the boundary conditions $u_n = \psi_n =
b_n  = 0 \forall n < 1$ or $\forall n > N$. We use a second-order
Adams-Bashforth method to solve the equations, a time step $\delta t
= 10^{-4}$ and $\nu = \eta = 10^{-8}$ in all our simulations. We
choose a Gaussian, stochastic forcing on the fourth shell ($n=4$) to
drive the system to a statistically steady state. Although in most
studies of shell models, a deterministic force is used, we chose a
stochastic forcing to make our shell model simulations consistent
with our DNS. A snapshot of the fluid kinetic energy spectrum,
obtained from our shell model studies, which
gives a good indication of the extent of the inertial range
obtained, are shown in Fig. \ref{Ekshell} with the K41 scaling indicated
by the thick black line. The extent of scaling, a little over 3 decades,
is typical of such shell models and which allows measurements of scaling exponents with
a higher degree of precision and confidence than in most DNS ~\cite{raynjp}. We show in Table 1 our equal-time scaling
exponents $\psi$ (column 2), $b$ (column 3) and $u$ (column 4)
fields; these exponents are calculated by using ESS, with respect to
the third-order structure functions, for 50 different statistically
independent statistically steady state configurations and quote the
mean of these as our exponents and the standard deviation as the
error-bars. We show the exponents for the velocity (red star) and the
concentration field (red filled-circle), in Fig.~\ref{DNS_zetap}  and for the gradient of the
concentration field (red star), in Fig. ~\ref{DNS_zetab}, as a function of $p$; it is clear
from the figures that there is clear multiscaling of the exponents
associated with $u$ (see Fig.~\ref{DNS_zetap} , red star) and $b$ (see Fig.~\ref{DNS_zetab} , red star) and that the two agree with each other
within error-bars (compare columns 3 and 4 in Table 1). In contrast,
the exponents for $\psi$ (see Fig.~\ref{DNS_zetap} , red filled-circle) shows simple scaling and is
indistinguishable, within error-bars, from the K41 prediction.
How much of our results from the shell model studies, as described
above, carries over to the actual Direct Numerical Simulations of
the SBF mixture? We now describe the results obtained from our
pseudospectral studies of the randomly forced symmetric binary fluid
mixture equations (Eqs.\ref{navierI} and \ref{advecI}) and compare
them with those obtained from the numerical solutions of our shell
models. In our DNS, we keep $y=4$, corresponding to K41 spectra for
the $\bf u$ and $\bf b$ fields, we use resolutions of $96^3$ and
$128^3$ with a cubic box of linear size $L=2\pi$, and periodic
boundary conditions. Our numerical scheme is identical to that in
Ref.~\cite{asain-prl}. We use hyperviscosity and hyperdiffusivity
together with ordinary viscosity and diffusivity. For the resolution
$128^3$, we are able to achieve Taylor microscale Reynolds number
$Re_\lambda\sim 150$. In Fig \ref{Ekdns}, we show a log-log plot of the fluid kinetic
energy spectrum obtained from our DNS. Although our Reynolds number is not very
high, we do obtain an inertial range close to three-quarters of a decade as can be seen
in the figure.  In the NESS obtained from these DNS we
calculate the exponents $\zeta_p^a$, by using ESS with respect to
the third-order structure function, from log-log plots of ${\mathcal
S}_p^a (r)$ versus ${\mathcal S}_3^a(r)$ ($a=u_i,b_i,\psi$). From
such plots, we use a modified local slope approach to obtain the
equal-time exponents : We calculate the exponents over various
ranges within the inertial range; we quote the mean as our exponent
and the standard deviation as the error-bar. We find that : (i) The
exponents $\zeta_p^u$ (Fig.~\ref{DNS_zetap}, blue triangle) and
$\zeta_p^b$ (Fig.~\ref{DNS_zetab}, blue triangle) display multiscaling very similar to
that in fluid turbulence:
$\zeta_2^m/\zeta_3^m>2/3,\,\zeta_p^m/\zeta_3^m<p/3,\,p>3,\,m=u,b$,
and, $\zeta_p^u/\zeta_3^u$ and $\zeta_p^b/\zeta_3^b$ are equal to
each other within our error-bars (compare columns 6 and 7 in Table
1); and (ii) $\zeta_p^\psi/\zeta_3^\psi\approx p/3$ (Fig.~\ref{DNS_zetap}, blue diamond). In addition, we
calculate the normalized probability distribution function (PDF)
$P[\delta a(r)]$ ($a=u_i,b_i,\psi$) for $r/\eta_d = 7.7$  (see Fig. ~\ref{PDF}). We find $P[\delta u(r)]$ and $P[\delta
b(r)]$ are nearly overlapping and have much longer tails than
$P[\delta\psi (r)]$. Furthermore, $P[\delta\psi (r)]$ is well
represented by a Gaussian of unit variance. In Fig.~\ref{prob2} we
show plots of $P[\delta\psi (r)]$ versus $\delta\psi (r)$ for three
different separations $r/\eta_d = 7.7,\,23.2,\,26.2$ in the inertial
range. A Gaussian of unit variance is again shown for comparison. We
find that for all values of $r$, the plots overlap with each other
and with the Gaussian. Also similar PDF plots for $u$ (Fig.~\ref{prob2u} and
$b$ (Fig.~\ref{prob2b}, for three different separations $r = /\eta_d = 7.7,\,23.2,\,26.2$
not only show a marked departure from a Gaussian (as indicated by a continuous dark
blue line) as was seen in Fig.~\ref{PDF}, but also no collapse of the curves for different $r$ (unlike
the case for $\psi$).
These PDFs further strengthen and provide compelling evidence  for  our main results
(i) and (ii) above. We present, in Table 1, the multiscaling
exponents $\zeta_p^\psi$ (column 5), $\psi_p^b$ (column 6), and
$\psi_p^u$ (column 7) from our DNS. We have checked that our results
from the two different resolutions for the DNS agree with each other
within error bars.

\begin{figure}[htb]
\includegraphics[height=6cm,width=6cm]{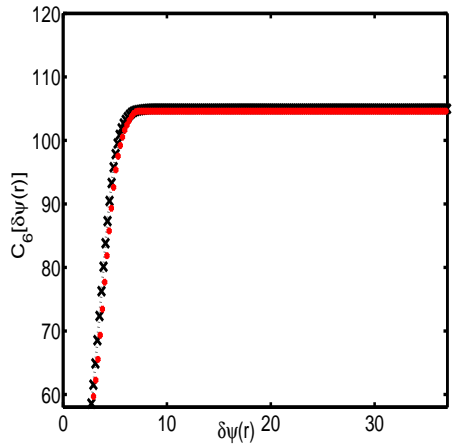} \caption{(Color
online) Convergence of the sixth-order accumulated moment $C_6[\delta\psi(r)]$
versus $\delta\psi(r)$ for two different separations $r=26.2\eta_d$ (black) and $7.7\eta_d$ (red)  in the inertial range. The two
curves overlap (see text).} \label{cumpsi}
\end{figure}

\begin{figure}[htb]
\includegraphics[height=6cm,width=6cm]{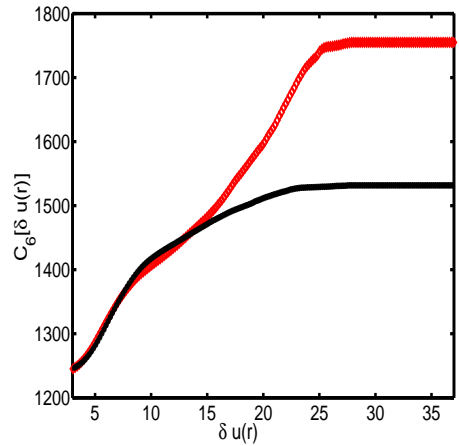} \caption{(Color
online) Convergence  of the sixth-order accumulated moment $C_6[\delta u(r)]$
versus $\delta u(r)$ for two different separations $r=26.2\eta_d$ (black, bottom) and $7.7\eta_d$ (red, top) in the inertial range. The two
curves do not overlap (see text).} \label{cumv}
\end{figure}

Given our modest resolution for our DNS, it is useful to examine how
far we are justified in calculating moments up to order 6. It is well
known that for higher order moments, the large contributions from the
tails of the PDFs make statistical convergence progressively poor.
In order to study statistical convergence, a good prescription is to
examine the convergence of the moments of the differences of the velocity and the
concentration fields ~\cite{gotoh}.
Hence we study the bulk contributions $C_6[\delta a]$ to the
sixth-order structure function (the highest order for which
we present results in this paper) $S_6^a(r)$ ($a=\psi,\,u)$ defined as:
$C_6(x) =\int_0^x x^6p(x)$ where $x = \delta\psi(r)$ or $\delta u$ (all suitably normalised),
and $p(x)$ is the PDF of the same, respectively. In
Fig.~\ref{cumpsi}, we show a plot of $C_6[\delta\psi(r)]$ versus
$\delta\psi(r)$ for two different $r=26.2\eta_d$ (black) and $7.7\eta_d$ (red): The
two curves overlap, as is expected for a Gaussian form for
$P[\delta\psi(r)]$ for various different values of $r$.
In Fig.~\ref{cumv} we show a
similar plot for $C_6^u[\delta u(r)]$ for the same two $r$ as before. Due to
the non-Gaussian nature of $P[\delta u(r)]$, the two curves for two different
$r$ {\em do not} overlap. These plots strongly display
statistical convergence of the corresponding sixth-order moments. We obtain similar
convergence for the gradient of the concentration field which we do not show here.

\section{Conclusion}

In summary, then, in this paper we have investigated the scaling and multiscaling properties
of a turbulent symmetric binary fluid mixture via detailed numerical simulations.
We find that ${\mathcal S}_p^{u,b}(r)$ exhibit
multiscaling similar to fluid turbulence in the inertial range, whereas
${\mathcal S}_p^\psi (r)$ exhibit simple $p/3$ K41-scaling
(within error bars). Moreover, the probability distributions
$P[\delta u(r)]$ and $P[\delta b(r)]$ are nearly overlapping and have
tails longer than that of $P[\delta\psi (r)]$ for $r$ in the inertial range.
We also propose a new shell model for the gradient of the concentration field and numerically
solve it as well as the shell model for scalar concentration field. Our results
from our shell models are in agreement with our DNS studies. Our results
are $3d$ analogues of those of Ref.~\cite{celani}, where simulations with particles in $2d$ flows
were used to show that the structure functions of the concentration field for the
SBF problem {\em do not} multiscale.
The results from both our DNS of the SBF equations and
shell-model studies are complementary to and agree well with each other and bring out the multiscaling of the
velocity and concentration gradient fields and simple scaling of the
concentration field. The validity of these conclusions is strengthened not only by the reliability of the
scaling ranges usually associated with the measurement of equal-time structure functions in shell models,
but also by the convincing PDFs that we obtain for various quantities in our DNS and the clear evidence of
statistical convergence which justifies measurements of equal-time exponents upto order 6 in our DNS.

Our results may be
explained from the analytical framework based on symmetry arguments developed in Ref.~\cite{abhik},
where it has been shown that  the presence of an additional continuous symmetry (kind of a gauge symmetry),
not present in the passive scalar turbulence model, is responsible for the simple scaling behaviour of
$\zeta_p^\psi$.
It would be interesting to investigate the properties of the turbulent NESS
of SBF at low temperature, below the consolute point, when instabilities leading to phase separation competes with turbulent mixing.
Work is in progress in this direction.  Finally, our results
may be tested in experiments similar to Ref.~\cite{war}.
One of the authors (AB) gratefully
acknowledges MPG(Germany)-DST(India) for partial financial support through
the Partner Group program.

\begin{table*}
\framebox{\begin{tabular}{c|c|c|c|c|c|c}
order$(p)$ & $\zeta^{\psi,{\rm shell}}_p$ & $\zeta^{b,{\rm shell}}_p$ & $\zeta^{u,{\rm shell}}_p$ & $\zeta^{\psi,{\rm DNS}}_p$ & $\zeta^{b,{\rm DNS}}_p$ & $\zeta^{u,{\rm DNS}}_p$ \\
\hline
 1 &  0.3334 $\pm$ 0.0001 & 0.3671 $\pm$ 0.0001  & 0.378 $\pm$ 0.005
    & 0.334 $\pm$ 0.001  & 0.372 $\pm$ 0.009 & 0.385 $\pm$ 0.009
   \\
 2 &  0.6660 $\pm$ 0.0009  & 0.698 $\pm$ 0.005 & 0.707 $\pm$ 0.007
   &0.677 $\pm$ 0.001  &  0.70 $\pm$ 0.01 & 0.710 $\pm$ 0.009
 \\
 3 &  1.0000  & 1.0000 & 1.0000
   & 1.000 & 1.000 & 1.000
  \\
 4 &  1.334 $\pm$ 0.002  &  1.277 $\pm$ 0.009 & 1.27 $\pm$ 0.01
 & 1.340 $\pm$ 0.002 & 1.280 $\pm$ 0.009  & 1.28 $\pm$ 0.01
\\

 5 &  1.665 $\pm$ 0.005 & 1.54 $\pm$ 0.02  & 1.51 $\pm$ 0.02
   & 1.671 $\pm$ 0.006  & 1.55$ \pm$ 0.01 & 1.52 $\pm$ 0.04
  \\
 6 &  1.995 $\pm$ 0.009 & 1.78 $\pm$ 0.03  & 1.75 $\pm$ 0.03
   &1.997 $\pm$ 0.008 &  1.78 $\pm$ 0.01 & 1.77 $\pm$ 0.06
   \\

\end{tabular}}
\caption{We show the various equal-time, order-$p$ exponents
obtained from our shell model (indicated by the superscript shell) and DNS (indicated by the superscript DNS) studies.
By comparing the
corresponding columns, we find an agreement between the exponents
obtained from DNS and the ones obtained from our shell models.}
\label{table}
\end{table*}


\begin{thebibliography}{99}
\bibitem{frisch}U. Frisch, {\em Turbulence: The Legacy of A.N. Kolmogorov}, Cambridge University Press, Cambridge (1995).
\bibitem{falcormp}G. Falkovich, K. Gawedzki and M. Vergassola,
Rev. Mod. Phys. {\bf 73}, 913 (2001).
\bibitem{rmp}P.C. Hohenberg and B.I. Halperin, Rev. Mod. Phys.
{\bf 49}, 435 (2004) and references therein.
\bibitem{chaikin}P.M. Chaikin and T.C. Lubensky, {\it Principles
of Condensed Matter Physics} (Cambridge University,
Cambridge, England, 2004).
\bibitem{K41a}A.N. Kolmogorov, Dokl. Akad. Nauk SSSR
{\bf 30}, 301 (1941).
\bibitem{K41b}A.N. Kolmogorov, Dokl. Akad. Nauk SSSR
{\bf 31}, 538 (1941).
\bibitem{kraich1}R. Kraichnan, Phys. Fluids {\bf 11}, 945 (1968).
\bibitem{kraich2}R. Kraichnan, Phys. Rev. Lett. {\bf 72}, 1016 (1994).
\bibitem{kraich3}R. Kraichnan, Phys. Rev. Lett. {\bf 78}, 4922 (1997).
\bibitem{obu}A. M. Obukhov, Izv. Akad. SSSR, Serv. Geogr. Geofiz. {\bf 13}, 58
 (1949).
\bibitem{corr}S. Corrsin, J. Appl. Phys. {\bf 22}, 469 (1951).
\bibitem{exp} H. L. Swinney {\em et al.}, {\em Phys. Rev. A}, {\bf 8}, 2586 (1973).
\bibitem{celani} A. Celani {\em et al}, {\em Phys. Rev. Lett.}, {\bf 89}, 234502 (2002).
\bibitem{abhik} A. Basu, {\em J. Stat. Mech.}, L09001 (2005).
\bibitem{ruiz} R. Ruiz and D. R. Nelson, {\em Phys. Rev. A}, {\bf 23}, 3224 (1981).
\bibitem{jkb1} M. K. Nandy {\em et al.}, {\em J. Phys. A}, {\bf 31}, 2621 (1998).
\bibitem{passive} M. Chertkov {\em et al}, {\em Phys. Rev. E}, {\bf 52}, 4924
(1995); M. Chertkov {\em et al}, {\em Phys. Rev. Lett.}, {\bf 76},
2706 (1996); K. Gawedzki and A. Kupiainen, {\em Phys. Rev. Lett.},
{\bf 75}, 3834 (1995); D. Bernard {\em et al.} {\em Phys. Rev. E},
{\bf 54}, 2564 (1996); L. Ts. Adzhemyan {\em et al.} {\em Phys.
Rev. E}, {\bf 58}, 1823 (1998).
\bibitem{mhd-basic} D. Montgomery, in {\em Lecture Notes on Turbulence}, edited by J. R. Herring and J. C. McWilliam (World Scientific,
Singapore, 1989); D. Biskamp, in {\em Nonlinear Magnetohydrodynamics}, edited by W. Grossman et al. (Cambridge University Press, Cambridge, England, 1993).
\bibitem{foot} In a renormalization group language fields $\bf b$
and $\bf B$ have the same canonical dimensions when the
respective equations of motion are driven by noises having variances
with same spatial scaling.
\bibitem{abprl} A. Basu {\em et al}, {\em Phys. Rev. Lett.}{\bf 81}, 2687 (1998).
\bibitem{yakhot} V. Yakhot and S.A. Orszag, {\em Phys. Rev. Lett.}, {\bf 57}, 1722 (1986); J.K. Bhattacharjee, {\em J. Phys. A}, {\bf 21}, L551 1988.
\bibitem{rgprobs} C.Y. Mou and Weichman, {\em Phys. Rev. Lett.}, {\bf 70}, 1101 (1993); G.L. Eyink, {\em Phys. Fluids}, {\bf 6}, 3063 (1994).
\bibitem{sl} Z. S. She and E. Leveque, {\em Phys. Rev. Lett.}, {\bf 72}, 336 (1994).
\bibitem{fluid-ess} R. Benzi {\em et al.}, {\em Phys. Rev. E}, {\bf 48}, R29 (1993); S. K. Dhar {\em et al}, {\em Phys. Rev. Lett.}, {\bf 78},
2964 (1997); S. Chakraborty {\em et al.}, {\em J. of Fluid Mech.}, {\bf 649}, 275 (2010).
\bibitem{goy} E. B. Gledzer, {\em Sov. Phys. Dokl.}, {\bf 18}, 216 (1973); K. Ohkitani and M. Yamada, {\em Prog. Theor. Phys.}, {\bf 81}, 329 (1989).
\bibitem{raynjp} S.S. Ray {\em et al.}, {\em New J. Phys.}, {\bf 10}, 033003 (2008).
\bibitem{jensen} M. H. Jensen and P. Olesen, {\em Physica D}, {\bf 111}, 243 (1998).
\bibitem{asain-prl} A. Sain {\em et al}, {\em Phys. Rev. Lett.}, {\bf 81}, 4377 (1998).
\bibitem{gotoh} T. Gotoh, D. Fukayama, and T. Nakano,  Phys. Fluids, {\bf 14}, 1065 (2002).
\bibitem{war} R. E. G. Poorte and A. Biesheuvel, {\em J. Fluid Mech.}, {\bf
461}, 127 (2002), A. Gylfason and Z. Warhaft, {\em Phys. Fluids}, {\bf 16},
4012 (2004), and references therein.
\end{thebibliography}
\end{document}